\documentclass[showpacs,aps,prl,twocolumn,amsmath,amssymb,epsfig]{revtex4-1}

\usepackage{graphicx}
\usepackage{epsfig}
\usepackage{dcolumn}% Align table columns on decimal point
\usepackage{bm}

\newcommand     {\beqa}[1]        { \begin{eqnarray} #1 \end{eqnarray} }

\begin{document}

\title{Temporal and spacial evolution of bursts in creep rupture}

\author{Zsuzsa Danku}
\author{Ferenc Kun}
\email{ferenc.kun@science.unideb.hu}
 \affiliation{Department of Theoretical Physics, University of Debrecen,
P.O. Box 5, H-4010 Debrecen, Hungary}

\begin{abstract}
We investigate the temporal 
and spacial evolution of single bursts and their statistics emerging in heterogeneous materials 
under a constant external load. Based on a fiber bundle model we demonstrate
that when the load redistribution is localized along a propagating crack front,
the average temporal shape of pulses has a right handed asymmetry, however, 
for long range interaction a symmetric shape with parabolic functional form is obtained. 
The pulse shape and spatial evolution of bursts proved to be correlated
which can be exploited in materials' testing. The probability distribution of the size and duration 
of bursts have power law behavior with a crossover to higher exponents 
as the load is lowered. The crossover 
emerges due to the competition of the slow and fast modes of local
breaking being dominant at low and high loads, respectively. 
 \end{abstract}

\maketitle

Crackling noise is a generic feature of a 
wide variety of slowly driven dynamic systems such as ferromagnetic materials, 
plastically deforming crystals, superconductors, fracture processes of 
heterogeneous materials and earthquakes 
\cite{sethna_crackling_2001,zapperi_nat2011,zapperi_nat2,dahmen_nat2011,zapperi_nat2012,alava_statmodfrac_2006}. 
Analyzing the time series of crackling events it was shown that crackling phenomena
exhibit a high degree of universality \cite{sethna_crackling_2001,dahmen_nat2011}: 
the size and duration of events, 
furthermore, the waiting times in between are characterized by power law distributions
with the same exponents in systems of different microscopic dynamics. 
Recently it has been demonstrated for Barkhausen 
noise that there are unique features of crackling which go beyond universality, i.e.\
the average temporal shape of single burst pulses proved to
provide direct information about the nature of correlations
in the microscopic dynamics 
\cite{zapperi_nat2011,zapperi_nat2,baldassarri_prl2003,sethna_pre_2002,mikko_pre_2006}.
For systems where the impulsive relaxation mechanism competes with slow ones,
a novel phase of crackling has been discovered very recently \cite{zapperi_nat2012}:
when the rate of external driving becomes comparable to the time scale of slow relaxation
large bursts emerge in a quasi-periodic manner which is accompanied by a crossover 
in the statistics of burst sizes and durations.

We present a theoretical investigation of crackling noise emerging 
during the creep rupture of heterogeneous materials focusing on single burst dynamics
in the presence of two competing failure mechanisms. 
Creep rupture has
a high technological importance for the safety of construction components 
and it is at the core of natural catastrophes such  as landslides, 
stone and snow avalanches, as well \cite{carpinteri_prl}. Crackling during creep 
is the consequence of the intermittent nucleation and propagation of cracks
which generate acoustic bursts. 
Despite the intensive research on rupture phenomena
\cite{dahmen_nat2011,alava_statmodfrac_2006,carpinteri_prl,maes_criticality,nechad_creep_2005,alava_jphysd_2009,vanel_jstat_2009,
maloy_prl2006,bonamy_prl2008,tallakstad_2011}, 
the temporal and spatial
evolution of single bursts, the effect of the competition of failure mechanisms
with different time scales, and their relevance for applications still 
remained an open fundamental problem. 

\begin{figure}
\begin{center}
\epsfig{bbllx=5,bblly=35,bburx=645,bbury=530,
file=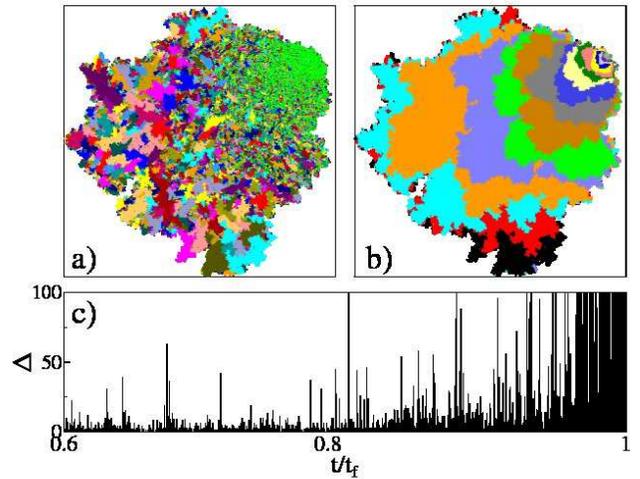, width=8.3cm}
  \caption{(Color online) $(a)$ Bursting activity during the creep rupture of a fiber bundle of size $L=401$.
Slowly damaging fibers (green) trigger bursts of immediate breaking (random
colors different from green). The crack started in the upper right corner of the figure. 
$(b)$ Advancement of the front of the same crack obtained such that
fibers broken in a time interval have the same color. $(c)$ Time series of bursts of $(a)$ 
as a stochastic point process. 
   \label{fig:crack}}
\end{center}
\end{figure}

To investigate the creep rupture of heterogeneous materials we use a generic fiber bundle 
model (FBM) introduced recently \cite{kun_damage_jstat,kun_prl2008_dam,halasz_2012}: 
the sample is discretized as a bundle of fibers on a square lattice 
of side length $L$. Fibers have a brittle 
response with identical Young modulus $E$. The bundle is 
subject to a constant external load $\sigma_0$ below the fracture strength $\sigma_c$
of the system parallel to the fibers. 
Fibers break due to two physical mechanisms:
immediate breaking occurs when the local load $\sigma_i$ on fibers exceeds 
their fracture strength 
$\sigma^{th}_{i}$. 
Time dependence is introduced such that those fibers, which remained intact, 
undergo an aging process accumulating damage $c_i(t)$. 
The damage mechanism represents the environmentally induced slowly developing
aging of materials e.g.\ thermally activated degradation
\cite{kun_damage_jstat,kun_prl2008_dam,halasz_2012}. 
The rate of damage accumulation $\Delta c_i$ is assumed 
to have a power law dependence on the local load 
$\Delta c_i = a\sigma_i^{\gamma}\Delta t,$
where $a=1$ is a constant and the exponent $\gamma$ controls the time scale of the 
aging process with $0\leq \gamma < +\infty$.  
Fibers can tolerate only a finite amount of 
damage and break when $c_i(t)$ exceeds the local damage threshold $c^{th}_i$.
Each breaking event is followed by a redistribution of load over the remaining intact
fibers. 
To capture the effect of the inhomogeneous stress field around cracks,
we assume localized load sharing (LLS), i.e.\ the load of broken fibers 
is equally redistributed over their intact nearest neighbors \cite{halasz_2012}.
\begin{figure}%[!h]
\begin{center}
\epsfig{bbllx=40,bblly=10,bburx=650,bbury=285,file=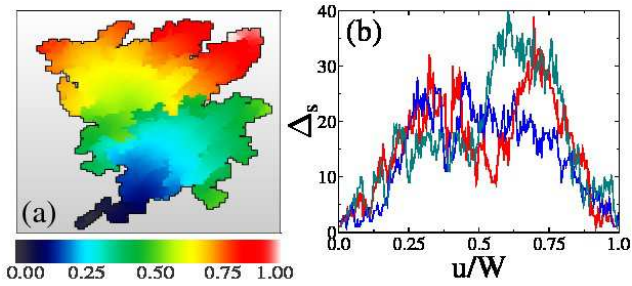, width=8.3cm}
\end{center}
  \caption{(Color online) $(a)$ Temporal and spatial evolution of a single burst of size $\Delta= 3485$. 
The color code represents the normalized time $0\leq u/W \leq 1$. The burst starts from
a small spot of broken fibers ({\it blue}) at the bottom left corner, then it gradually expands 
({\it green, yellow, red}) and finally stops again in a small spot 
({\it white}) at the top right corner. $(b)$ The size of sub-bursts $\Delta_s$ as a function
of $u/W$ for a fixed duration $W=253$. The red curve corresponds to the burst of $(a)$.
  \label{fig:crack_pulse}
}
\end{figure}
The structural disorder of the material is represented by 
the randomness of breaking thresholds $\sigma^{th}_i, c^{th}_i$, $i=1,\ldots , N=L^2$. 
We assume that both thresholds are uniformly distributed in an 
interval $[1-\delta,1+\delta]$, where $\delta=1$ (high disorder) for $\sigma^{th}$.
To promote the effect of stress concentrations, 
lower disorder $\delta=0.3$ is considered for damage $c^{th}$, 
while the exponent $\gamma$ of the 
damage law is set to a high value $\gamma=5$ \cite{halasz_2012}. 
%Our model has proven very successful in reproducing measured creep behavior
%\cite{kun_damage_jstat,kun_prl2008_dam,halasz_2012}.
In the following, simulation results will be presented for the lattice size $L=401$,
larger system sizes up to $L=1201$ are considered for finite size scaling analysis. 

Figure \ref{fig:crack}$(a)$ presents an example of the time evolution 
of a crack in our FBM under the load $\sigma_0/\sigma_c=0.01$.
The separation of time
scales of the slow damage process and of immediate breaking leads to a highly complex 
time evolution in agreement with experiments \cite{kun_damage_jstat,kun_prl2008_dam}: 
after the crack nucleated, fibers
mostly break due to slow damaging and generate an advancing crack front where the stress of broken 
fibers is concentrated. Beyond a certain 
crack size the subsequent load increments become 
sufficient to trigger bursts of immediate breakings locally accelerating the front. 
As a consequence, the time evolution of creep rupture
occurs as a series of bursts 
separated by silent periods of slow damaging. 
The size of bursts $\Delta$ is defined as the number
of fibers breaking in a correlated trail.  
For clarity, in Fig.\ \ref{fig:crack}$(b)$ the crack 
front is presented at several times, while
Fig.\ \ref{fig:crack}$(c)$ shows the time series of bursts, i.e.\
the burst size $\Delta$ as a function of time $t$ normalized 
with the lifetime $t_f$ of the system 
\cite{nechad_creep_2005,
alava_jphysd_2009,vanel_jstat_2009,kun_prl2008_dam,halasz_2012}. 
See Supplemental Material at [URL] for an animation of the bursty crack growth. 

\begin{figure}
\begin{center}
\epsfig{bbllx=65,bblly=70,bburx=760,bbury=640,
file=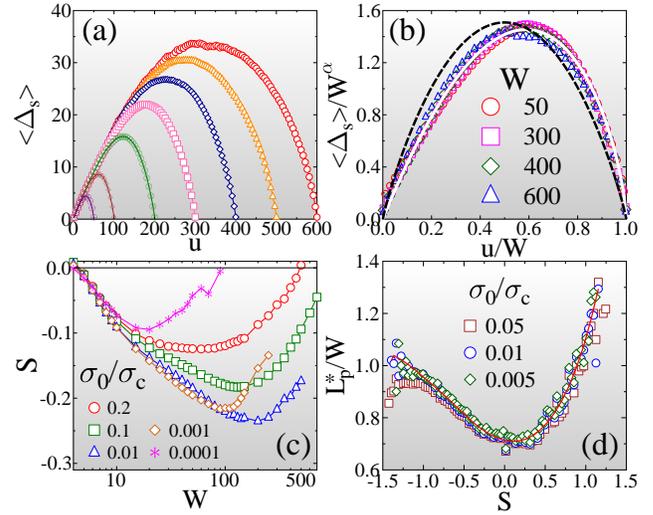, width=8.1cm}
\end{center}
  \caption{(Color online) $(a)$ Average pulse shapes for durations $W=50, 100, 200, 300, 400, 500, 600$
for LLS FBMs obtained at the load $\sigma_0/\sigma_c=0.01$. 
$(b)$ Scaling collapse of average pulse shapes of different duration for LLS. 
The white continuous line represents the fit with the scaling function $f(x)$.
The dashed black line presents the scaling function of the mean field limit of FBM for comparison. 
$(c)$ Skewness $S$ of pulse shapes as a function of the 
duration $W$ at different loads.
$(d)$ The number of perimeter sites of bursts touching the crack
front $L_p^*$ divided by the duration $W$ as a function of the 
corresponding skewness of bursts.
  \label{fig:pulse_all_lls}}
\end{figure}
Single bursts typically start with the immediate breaking of a few fibers. 
Due to load redistribution, additional breakings 
are triggered so that bursts 
gradually evolve through sub-avalanches and stop when all the intact fibers 
along the burst boundary can sustain the local load. This avalanche dynamics
is controlled by the range of interaction so that similar evolution would arise
under quasi-statically increasing external load, as well 
\cite{kun_damage_jstat,kun_prl2008_dam,halasz_2012,hansen_crossover_prl}.
Figure \ref{fig:crack_pulse} presents for a burst of size $\Delta = 3485$ that the outbreak
starts from a small localized spot which then gradually expands to a broad region 
followed by the subsequent reduction of the breaking activity. 
The temporal profile of bursts can be characterized 
by recording the size $\Delta_s$ of sub-avalanches 
as a function of the internal time step $u$ over the 
duration $W$ of the burst, where $1\leq u \leq W$ holds.
Comparing $\Delta_s(u)$ curves of bursts of the same duration $W= 253$ 
in Fig.\ \ref{fig:crack_pulse}$(b)$ the stochastic nature of avalanche 
dynamics is apparent.  

The average pulse shape $\left<\Delta_s(u,W) \right>$ of bursts is presented
in Fig.\ \ref{fig:pulse_all_lls}$(a)$ as a function of
time $u$ varying the pulse duration $W$.
The $\left<\Delta_s(u,W) \right>$ curves have a 
right-handed asymmetry, i.e.\ they can be described by a nearly parabolic 
shape where the maximum 
of the inverted parabola is shifted from the middle ($W/2$) to higher values.
Figure \ref{fig:pulse_all_lls}$(b)$ demonstrates that rescaling 
$\left<\Delta_s(u,W) \right>$ with an appropriate power $\alpha$ of the duration $W$,
the pulse shapes of different $W$ can be collapsed on
a master curve as a function of the normalized time $x=u/W$. 
The good quality data collapse implies the scaling form
\begin{eqnarray}
\left<\Delta_s(u,W) \right> = W^{\alpha} f(u/W), 
\label{eq:scaling_pulse}
\end{eqnarray}
where both the scaling function $f(x)$ and the scaling exponent $\alpha$
encode information about the jerky crack propagation. 
The right-handed asymmetry of the scaling function 
$f(x)$ shows that bursts start slowly, then gradually 
accelerate, and finally stop suddenly as the front gets pinned. 
The average pulse shape can be described by the function
$f(x) = Ax(1-x)^{\beta}$,
where $A$ determines the initial acceleration and the exponent $\beta < 1$ describes
the observed right-handed asymmetry. In Fig.\ \ref{fig:pulse_all_lls}$(b)$ 
collapse was achieved with $\alpha=0.7$, while the fit of the scaling function 
was obtained with $A=4.65$ and $\beta = 0.65$. 

In order to understand 
the role of the range of load sharing in shaping temporal pulses, we analyzed the mean
field limit of our FBM \cite{kun_damage_jstat,kun_prl2008_dam,
halasz_2012}. In this case the load of broken fibers
is equally shared by all the remaining intact ones so that no stress concentration,
and hence, no spatial correlation  can arise
in the bundle \cite{hansen_crossover_prl}. 
Fig.\ \ref{fig:pulse_all_lls}$(b)$
presents the scaling function $f(x)$ of the mean field simulations of a bundle
of $N=10^7$ fibers, where the symmetric parabolic shape is evidenced
with the scaling exponent $\alpha=1$, similarly to mean field avalanches 
of slip events in plastically deforming solids, and of particle
rearrangements in sheared granular matter 
\cite{zapperi_nat2011,zapperi_nat2,dahmen_nat2011,baldassarri_prl2003}.

To quantify the degree of asymmetry of pulse shapes we 
calculated the skewness $S$ as a function of $W$, where $S$ is defined
as the ratio of the third cumulant and the $3/2$ power of the second
cumulant of the pulse \cite{zapperi_nat2011}.
% \begin{eqnarray}
% S(W)=\frac{\int_0^W du (u-\overline{u})^3 \left<\Delta_s(u,W)\right>}
% {\left[\int_0^W du (u-\overline{u})^2 \left<\Delta_s(u,W)\right>\right]^{3/2}},
% \end{eqnarray}
% where $\overline{u}=(1/W)\int_0^W\left<\Delta_s(u,W)\right>u du$ denotes the mean of $u$
% \cite{zapperi_nat2011}.
In Fig.\ \ref{fig:pulse_all_lls}$(c)$ the skewness $S$ of pulse shapes 
is negative in agreement with the observed right-handed asymmetry,
however, the value of $S$ has a strong dependence on the pulse duration $W$: 
short pulses are flat and symmetric, hence, $S\approx 0$ follows in this range. 
Bursts of high duration tend again to be symmetric, since as they evolve the structural disorder
of the material becomes dominating, which favors symmetric 
pulse shapes \cite{zapperi_nat2011,zapperi_nat2,dahmen_nat2011,baldassarri_prl2003}. 
The most remarkable feature is that at each load a characteristic 
time scale $W_{max}$ emerges where the degree of asymmetry $|S|$ has a maximum.  
Both the characteristic 
duration $W_{max}$ and strength of asymmetry $|S|_{max}$ of pulses have a
strong dependence on the external load $\sigma_0/\sigma_c$, and
they reach a maximum nearly at the same load $\sigma_0/\sigma_c\approx 0.01$.
\begin{figure}
\begin{center}
\epsfig{bbllx=30,bblly=20,bburx=700,bbury=330,file=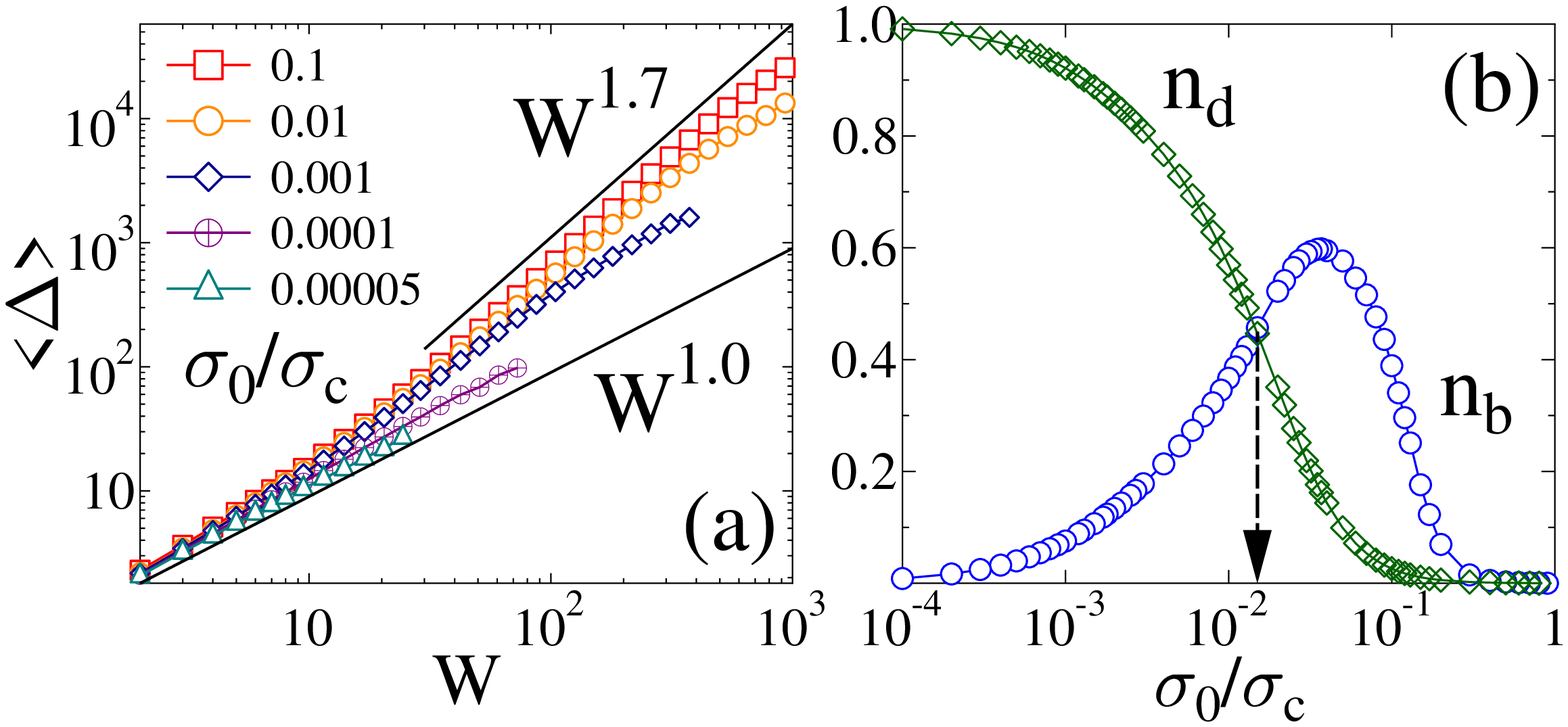, width=8.4cm}
\end{center}
  \caption{ 
(Color online) $(a)$ Average size of bursts $\left<\Delta\right>$ as a function 
of the duration $W$ for several load values $\sigma_0$. 
$(b)$ The fraction of fibers breaking due to
damage $n_d$ and to avalanches $n_b$ as function of $\sigma_0$.
  \label{fig:averages}}
\end{figure}

After starting from a localized spot, an avalanche has more chance
to advance if sub-bursts involve overloaded fibers at the front, as well, instead of
just propagating forward ahead of the crack front (see Fig.\ \ref{fig:crack}$(a,b)$). 
Consequently, the temporal profile and the geometry of avalanches with respect to the local position along
the crack front get correlated. To quantify this correlation
we measured the number of those perimeter sites $L_p^*$ of bursts which are located 
at the crack front. In Fig.\ \ref{fig:pulse_all_lls}$(d)$ the value of the 
ratio $L_p^*/W$ is plotted as a function
of the skewness $S$ of the corresponding pulse. For symmetric pulses 
($S\approx0$) the ratio $L_p^*/W$ has a low value which shows that these bursts 
were mainly moving forward 
where structural disorder dominates. However, for asymmetric pulses  
$L_p^*/W$ increases so that these avalanches involve a large fraction 
of overloaded fibers at the crack front which accelerates spreading. 
Our result has the important consequence that by measuring pulse shapes 
one can infer the spatial propagation of bursts. 

\begin{figure}
\begin{center}
\epsfig{bbllx=20,bblly=15,bburx=730,bbury=600,file=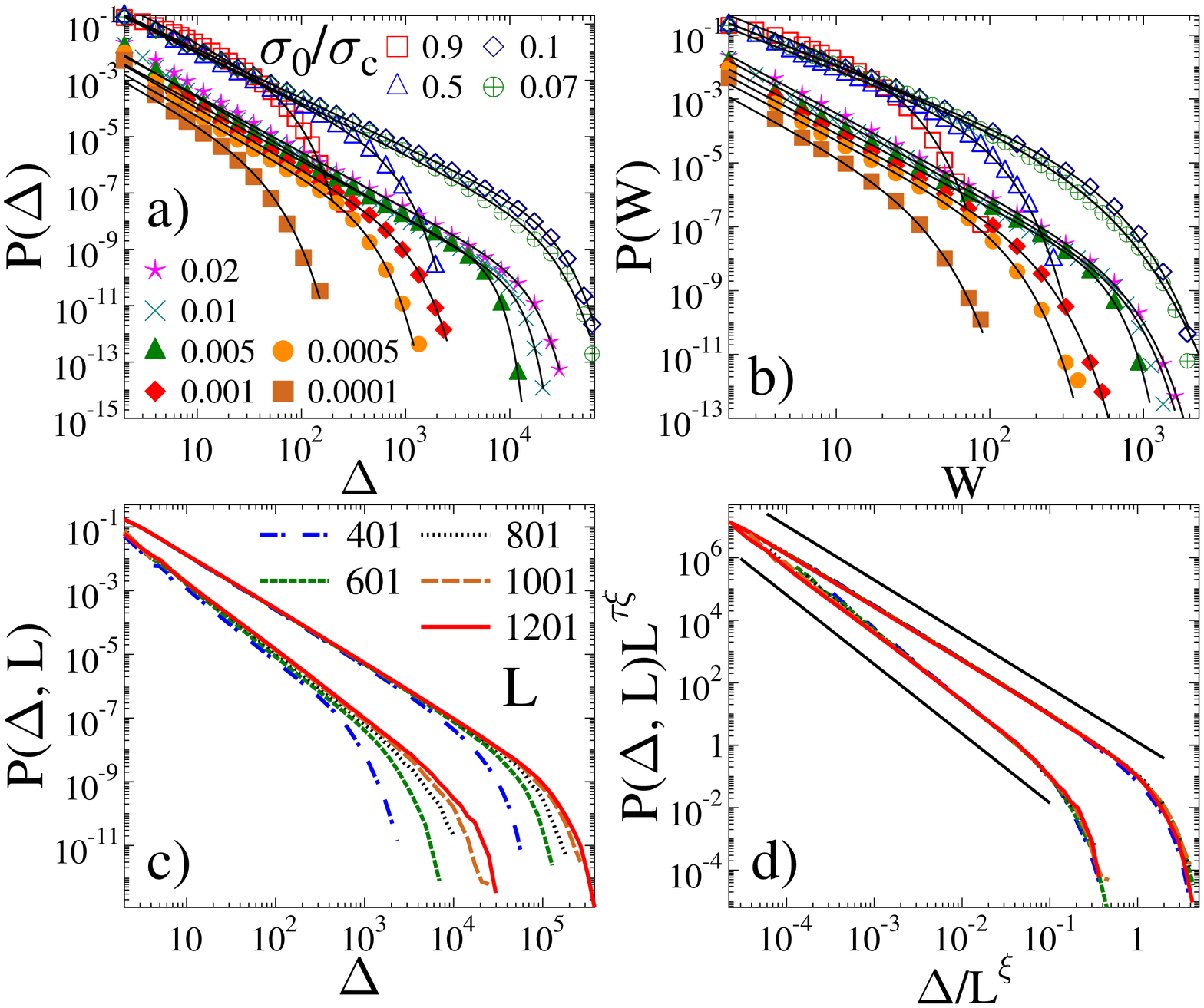, width=8.1cm}
\end{center}
  \caption{(Color online) Probability distribution of burst size $p(\Delta)$ $(a)$ and duration $p(W)$ $(b)$ 
varying the external load $\sigma_0$. The lines are fits 
with Eqs.\ (\ref{eq:burst_statistics_delta}). 
$(c)$ Burst size distributions for different system sizes $L$ obtained at two load values $\sigma_0/\sigma_c=0.1$ and $0.001$
for the upper and lower groups of curves, respectively.
$(d)$ High quality collapse of the distributions of $(c)$ obtained by rescaling 
with $L$ using the scaling exponents $\xi=1.6$, $\tau=1.75$ and $\xi=2.4$, $\tau=2.4$ for the lower and upper
curves, respectively. The straight lines represent power laws of slope $\tau$. 
In $(a)$ and $(b)$ the vertical axis, while in $(d)$ both axis are arbitrarily rescaled to better see
the data.
  \label{fig:distrib}}
\end{figure}

Eq.\ (\ref{eq:scaling_pulse}) implies that the average burst size 
$\left<\Delta\right>$ must scale with the duration as
$\left<\Delta\right> \sim W^{1+\alpha}$.
To verify the scaling structure we determined 
the function $\left<\Delta\right>(W)$ numerically 
by directly averaging the size of bursts $\Delta$ at fixed 
durations $W$. It can be observed in Fig.\ 
\ref{fig:averages}$(a)$ that the simulation results agree very well with
the analytic prediction, however, the asymptotic value of the power law exponent 
depends on the external load: at low loads the average burst size proved to be proportional
to the duration $\left<\Delta\right>\sim W$ 
with $\alpha=0$. Consequently, in this load range the pulse shape is not parabolic 
but instead it has a flat, symmetric functional form \cite{maloy_prl2006,bonamy_prl2008,tallakstad_2011} 
in agreement with the 
skewness value $S\approx 0$ (see also Fig.\ \ref{fig:pulse_all_lls}$(c)$). 
During the creep process the external load controls the relative 
importance of slow damaging and immediate breaking. At high load $\sigma_0\to \sigma_c$ already a few
damage breakings are sufficient to trigger extended bursts so that their size increases faster 
with the duration characterized by a higher exponent  $\left<\Delta\right>\sim W^{1.7}$
with $\alpha=0.7$ (see Fig.\ \ref{fig:averages}$(a)$). This crossover 
has also important consequences for the statistics of 
the size and duration of bursts. In Figs.\ \ref{fig:distrib}$(a,b)$ 
the probability distribution of the burst size $p(\Delta)$ and duration $p(W)$ 
have power law functional form at all load values followed by a cutoff
\beqa{
  p(\Delta) = \Delta^{-\tau} g(\Delta/\Delta_0),\ \ \ p(W) = W^{-z} h(W/W_0), \label{eq:burst_statistics_delta}
} 
where the functions $g(x)$ and $h(x)$ can be well approximated 
by exponentials. A very important feature of the results is that not only the cutoffs
$\Delta_0$ and $W_0$ depend on the external load $\sigma_0$, but the distributions
exhibit a crossover between two regimes of different exponents. The crossover point 
$\sigma_0^*/\sigma_c = 0.025(4)$ falls close to the load where bursts reach the 
highest asymmetry (compare to Fig.\ \ref{fig:pulse_all_lls}$(c)$). 
It is important to note that the crossover emerges due to the competition of the slow and fast
breaking modes of fibers as the external load is varied, in a close analogy to the mechanism which leads
to the self-organized avalanche oscillator in crystal plasticity \cite{zapperi_nat2012}.
At high loads $\sigma_0>\sigma_0^*$ large avalanches are triggered with longer durations, however,
when approaching the critical load $\sigma_c$ it gets more likely that once an avalanche 
started it cannot stop and leads to catastrophic failure. Consequently, the power law exponents 
$\tau$ and $z$ have relatively low values $\tau=1.75$ and $z=1.82$, and the 
cutoffs $\Delta_0$ and $W_0$ decrease with increasing load
$\sigma_0$ (see Figs.\ \ref{fig:distrib}$(a,b)$). On the contrary, low load $\sigma_0<\sigma_0^*$ 
favors the damage breaking, i.e.\ 
long damage sequences are needed to trigger bursts which have rather limited sizes.
Hence, this regime is characterized by higher exponents $\tau=2.4$ and $z=2.55$ 
of the distributions $P(\Delta)$ and
$P(W)$ and the cutoffs $\Delta_0$ and $W_0$ increase when approaching the crossover load from below.
To quantify the relative importance of the slow and fast failure modes we determined 
the fraction of fibers breaking due to damage $n_d$ and in avalanches $n_b$ as
function of load. It can be seen in Fig.\ \ref{fig:averages}$(b)$ that the fraction $n_d$ 
of damage breakings monotonically decreases with $\sigma_0$, while $n_b$ first increases 
and has a maximum at $\sigma_0/\sigma_c=0.045(6)$. Note that the two 
fractions become equal $n_d=n_b$ at the load  $\sigma_0/\sigma_c=0.018(5)$ which
falls very close to the crossover point supporting the above arguments.
To analyze the finite size dependence of the distributions $p(\Delta)$ and $p(W)$ we
carried out simulations at fixed loads for several lattice sizes 
$L=401, 601, 801, 1001, 1201$. As representative examples, 
Fig.\ \ref{fig:distrib}$(c)$ presents for two load values $\sigma_0/\sigma_c=0.1$ and $0.001$
that the cutoff burst size of $p(\Delta,L)$ increases with the system size $L$. 
Rescaling the two axis with appropriate powers of $L$ a high quality data collapse is
obtained in Fig.\ \ref{fig:distrib}$(d)$ which implies the scaling structure 
$p(\Delta,L)=\Delta^{-\tau}\phi(\Delta/L^{\xi})$. The cutoff exponent $\xi$ has the values
$2.4$, $1.3$ ($\sigma_0<\sigma^*$) and $1.6$, $1.0$ ($\sigma_0>\sigma^*$), 
for the distribution of burst sizes and durations, respectively. 
The analysis demonstrates the robustness of the critical exponents $\tau$ and 
$z$ of the system.

In conclusion, investigating the dynamics of single bursts 
in FBMs we revealed a rich spectrum of novel aspects of rupture processes. The average 
shape of burst pulses proved to be sensitive to the range of 
load redistribution. Since the evolution of bursts is controlled 
by the overloads at the avalanche frontier, we revealed that from the temporal pulse
shape one can infer the spatial advancement of bursts.  
The statistics of burst sizes and durations is characterized by power law distributions,
however, the competition of the failure modes with widely separated time scales
leads to the emergence of a crossover between two regimes of different exponents: approaching
the critical load large bursts are triggered implying low exponents for the distributions,
while below a characteristic load slow damaging dominates giving rise to higher exponents.
A very interesting future extension of our study is to perform tests
where $\sigma_0$ is increased at a constant rate $r$. The current 
calculations are at the limiting case $r=0$. Under such conditions 
{\it avalanche precursors} are expected in the timeseries as $r$ increases,
namely, quasi-periodic large avalanches, before rupture, as predicted
and observed in the modeling/experimental approach of Ref.\ \cite{zapperi_nat2012} in the
context of crystal plasticity and consistently observed in granular
experiments of avalanches in jammed granular matter at an incline \cite{granular}. 
The quasi-periods of events emerge due to the dynamic competition between the
two timescales of the external driving and the intrinsic aging \cite{zapperi_nat2012}.

\begin{acknowledgments}
We thank the projects TAMOP-4.2.2.A-11/1/KONV-2012-0036, 
TAMOP-4.2.2/B-10/1-2010-0024, TAMOP 4.2.4.A/2-11-1-2012-0001, OTKA K84157, 
and ERANET\_HU\_09-1-2011-0002.
\end{acknowledgments}

\end{document}